\documentclass[prb,preprintnumbers,amsmath,amssymb]{revtex4}

\usepackage{graphicx}
\usepackage{dcolumn}
\usepackage{bm}

\begin{document}

\title {Tunable steady-state domain wall oscillator with perpendicular magnetic anisotropy}

\author{A. Bisig}
\author{O. Boulle}
\author{ M. Kl\"aui}
\thanks{Also at: Zukunftskolleg, Universität Konstanz, 78457 Konstanz, Germany; Electronic Mail: mathias@klaeui.de}
\affiliation{Fachbereich Physik, Universit\"at Konstanz, Universit\"atsstra\ss e 10, 78457 Konstanz, Germany}



\date{\today}

\begin{abstract} 

We theoretically study domain wall oscillations upon the injection of a dc current through a geometrically constrained wire with perpendicular magnetic anisotropy. The oscillation frequency spectrum can be tuned by the injected current density, but additionally by the application of an external magnetic field independent of the power. The results of analytical calculations are supported by micromagnetic simulations based on the Landau-Lifshitz-Gilbert equation. The simple concept of our localized steady-state oscillator might prove useful as a nanoscale microwave generator with possible applications in telecommunication or for rf-assisted writing in magnetic hard drives.

\end{abstract}


\maketitle


The recent discovery that a spin polarized current can exert a torque on a magnetization through transfer of spin angular momentum has opened a new way to manipulate magnetization.\cite{stilesmiltatbook} Spin-polarized currents can be used to reverse a magnetization in multilayer pillar elements \cite{katine2000} and interact with domain walls leading to current induced domain wall motion in the direction of the electron flow.\cite{yamaguchi2004,klaui2005} In the nanopillar geometry the spin transfer torque can compensate for the magnetic damping and this leads to sustained magnetization precession dynamics that are converted into microwave emission by magnetoresistive effects. \cite{kiselev2003}  Recently new schemes based on the oscillations of magnetic domain walls due to spin-polarized currents have been studied.\cite{zhang2007,ono2008,matsushita2009,franchin2008} These schemes led to the concept of spin torque oscillator (STO) based on domain walls, which may prove useful as for applications in telecommunication or for rf-assisted writing in magnetic hard drives.

Microwave generation due to the small angle precession of a magnetic free layer in nanopillar structures\cite{kiselev2003} leads to small output power, so here the challenge is to increase the power of the STO. In complicated nanopillar multilayer structures, where an out-of-plane magnetic injector layer is used to polarize the current, full angle precessions of a magnetic free layer in combination with a magnetic tunnel junction gives rise to higher output power.\cite{houssamedine2007} However a simpler and easier to fabricate system of magnetization exhibiting full angle precessions upon current injection is a oscillating magnetic domain wall. Here the question is, whether sustained precession of a pinned domain wall induced by spin transfer can be obtained. The spin transfer torque acts in the case of domain walls differently from anti-damping. This was recognized early on by L.~Berger in a seminal paper\cite{berger1986} and recently it was shown that domain wall oscillation can be obtained through the Walker precession phenomena where the whole domain wall structure oscillates periodically at microwave frequencies.\cite{zhang2007} But the large current density necessary to attain this Walker regime has so far always led to domain wall motion in the standard wire geometry in soft magnetic materials, making it impossible to pin the domain wall during tho oscillation. 

Several schemes have been proposed to solve this problem, including the use of wires with an artificial local gradient of the magnetic damping \cite{zhang2007} or extremely narrow wires with nanoscale lateral dimensions\cite{ono2008} resulting in lower critical current for Walker precession, but these approaches are not very realistic for the design of future devices. For three dimensional geometrically confined domain walls which are complicated to fabricate, inside a magnetic bridge between two electrodes a similar behaviour is presented. \cite{matsushita2009} While the concept of a domain wall oscillator seems appealing, a simple and more realistic system is missing. Furthermore in the approaches put forward so far, there is no possibility to tune the frequency independently of the output power, which is a key prerequisite for a device.

In this letter, we show that in perpendicularly magnetized materials, Walker precession of a pinned domain wall can be easily obtained by pinning the domain wall in a simple notch geometry. By properly choosing the constriction geometry dimensions, a small domain wall demagnetizing field can be attained, while keeping the domain wall strongly pinned. The steady state oscillations are described by analytical equations based on the Landau-Lifshitz-Gilbert (LLG) equation with spin torque-terms. Micromagnetic simulations are then performed and show that precession occurs at zero applied field and is associated with large angle oscillation of the domain wall magnetization. The oscillation frequency depends linearly on the injected current and can be tuned over a wide range. In addition, higher order modes are revealed which can be controlled by a small transverse magnetic field. These results open a new route for a novel kind of a spin transfer oscillator operating at zero field with a simple geometry and with potential for large output power in more sophisticated implementations.

First we consider a one dimensional perpendicular system, where the magnetization $\mathbf{M}$ is turning from $M_z/\lvert \mathbf{M} \rvert=1$ to $M_z/\lvert \mathbf{M}\rvert=-1$, see Fig. 1. We describe this system with an analytical model, based on the Landau-Lifshitz-Gilbert equation with spin torque terms: \cite{li2004,zhang2004}
\begin{equation}
\frac{\partial \mathbf{m}}{\partial t} = -\gamma_{0}\mathbf{m}\times\mathbf{H}_{eff} + \alpha\mathbf{m}\times\frac{\partial\mathbf{m}}{\partial t}-\left(\mathbf{u}\cdot\nabla\right)\mathbf{m}+\beta\mathbf{m}\times\left[\left(\mathbf{u}_{S}\cdot\nabla\right)\mathbf{m}\right],
\end{equation} 
where $\mathbf{m}=\frac{\mathbf{M}}{M_{S}}$ is the unit vector along the local magnetization direction, $\mathbf{H}_{eff}$ is the effective magnetic field including the external field, the anisotropy field, the magnetostatic field and the exchange field. The spin current drift velocity $\mathbf{u}=\frac{\mathbf{j}_{e}P\mu_{B}}{eM_{S}}$ describes the spin current associated with the electric current in a ferromagnet, where $P$ is the spin polarization of the current, $\mu_{B}$ is the Bohr magneton, $e$ is the (positive) electron charge and $\mathbf{j}_{e}$ is the current density.\cite{thiaville2005} The dimensionless constant $\beta$ describes the degree of nonadiabacity between the spin of the non-equilibrium conduction electrons and the local magnetization.

For the magnetization dynamics of this system, the pinned domain wall can be described by two collective coordinates; the domain wall center position $q(t)$ and the domain wall tilting angle $\psi(t)$. Following the approach of Jung \cite{jung2008} we obtain two equations, describing our system, referred to as the one dimensional collective coordinates model:
\begin{eqnarray}
     \Delta_{0}\dot\psi -\alpha\dot q & = & \beta u + \frac{\gamma_{0}\Delta_{0}}{2M_{S}}\left( \frac{\partial\epsilon}{\partial q}\right), \label{eq:2}\\
     \dot q + \alpha\Delta_{0}\dot\psi & = & -u - \frac{\gamma_{0}\Delta_{0}}{M_{S}}K_{d}\sin 2\psi, \label{eq:3}
\end{eqnarray}
where $\Delta_{0}$ is the constant domain wall width, $K_{d}$ is the effective domain wall anisotropy and $\epsilon(q)$ is the domain wall potential energy per unit cross-sectional area, representing the pinning potential of the wire constriction geometry.\cite{thomas2006} 

In the case where the effective wall anisotropy $K_{d}=K_{y}-K_{x}$ vanishes, corresponding to a geometry where the Bloch and N\'eel domain wall have the same magnetostatic energy and no energy barrier in between, equations (\ref{eq:2}) and (\ref{eq:3}) lead to a steady state oscillation of the polar angle $\psi(t)$:
\begin{equation}
\label{eq:4}
\dot{\psi}=-\frac{u}{\alpha\Delta_{0}}.
\end{equation}
This in-plane rotation corresponds to a continuous and repeated transition from Bloch to N\'eel wall and vice versa (Fig. 1). Note that in the adiabatic limit there is no need for a pinning potential in order to permit steady state domain wall oscillation when the magnetostatic energy difference $K_d$ vanishes.

However in realistic wire geometries, inhomogeneous demagnetization fields lead to energy barriers between the Bloch and N\'eel wall and the effective wall anisotropy never vanishes completely. Even by proper tuning of the constriction width $w$ and wire thickness $t$, the magnetostatic energy difference between Bloch and N\'eel wall cannot be reduced absolutely to zero, due to small asymmetric stray fields originating from the wire geometry. Hence for $\lvert K_{d}\rvert > 0$ the domain wall starts to rotate, when the current density $j_{e}$ is larger than a certain threshold value $J_{c}$, given by\cite{tatara2004}
\begin{equation}
\label{eq:5}
J_{c} = \frac{e\gamma_{0}\Delta_{0}}{P\mu_{B}}\lvert K_{d} \rvert.
\end{equation}
This critical current density represents a minimal spin-torque, which has to be applied in order to turn the magnetization from a N\'eel to Bloch wall. So here we need an additional pinning potential in order to pin the domain wall center position. Note that both, the linear dependence of the oscillation frequency of the injected current density Eq. (\ref{eq:4}), as well as the critical current density $J_c$ are independent of the non-adiabatic spin-torque constant $\beta$.

We consider a system consisting of a ferromagnetic structure with perpendicular magnetic anisotropy with a geometrical confinement as shown in Fig. 1. The outer dimensions of the wire are $500\times60\times7~\text{nm}^{3}$ and for this material $K_d$ is minimal at a constriction width $w$ of $16~\text{nm}$. Note that the constriction with $w$ depends strongly on the material parameters an can be easily made larger for other out-of-plane materials. Here we assume typical material parameters for $\mathrm{Co}/\mathrm{Pt}$ multilayer with perpendicular magnetic anisotropy:\cite{boulle2008,metaxas2007} $M_{S}  = 1.4 \cdot 10^{6}~\text{A}/\text{m} $, $A = 1.6 \cdot 10^{-11}~\text{J}/\text{m}$, the effective perpendicular magnetic anisotropy $K_{eff}=2.7 \cdot 10^{5}~\text{J}/\text{m}^{3}$ and Gilbert damping parameter $\alpha = 0.15$. 

The results of our micromagnetic simulations employing LLG micromagnetics Simulator \cite{scheinfein} are shown in Fig. 2, illustrating the linear dependence of the domain wall oscillation frequency on the injected current density $j_{e}$ for two values of $\alpha$. For the simulation, the system is divided in a rectangular mesh with finite elements of $2\times2\times7~\text{nm}^{3}$, smaller than the exchange length for $\mathrm{Co}/\mathrm{Pt}$ which is $l_{ex}=3.6~\text{nm}$. The linear behaviour is in agreement with the simple one dimensional model and can be observed over a broad frequency range between $500~\text{GHz}$ and $3.5~\text{GHz}$. The critical current density $J_{c}$ is marked as red points, see Fig. 2, and it is not dependent on $\alpha$ and calculated as $J_{c}^{sim}= 1.34\cdot10^{11}~\text{A}/\text{m}^{2}$ (corresponding to spin drift velocity $u=1.11~\text{m/s}$). The theoretical value $J_{c}=5.7\cdot10^{9}~\text{A}/\text{m}^{2}$ given by Eq. (\ref{eq:5}), assuming a constant domain wall width $\Delta_{0}=\sqrt{A/K_{eff}}=3.2~\text{nm}$, is lower than the value extracted from our simulations $J_{c}>J_{c}^{sim}$. This is due to the fact that, the potential landscape for the polar angle $\psi$ is modulated by the  the wire constriction geometry in a way, that additional spin-torque has to be applied, in order to turn the magnetization inside the domain wall and this is not taken into account in the one dimensional calculations.

The nonlinearity at higher current densities can be explained by deformations exhibited by the the domain wall just before depinning, see Fig. 2 inset \textbf{A,B}. For a low current density $j_e=1.79\cdot10^{11}~\text{A}/\text{m}^2$ the domain wall center position $q$ is not shifted $q=(0\pm0.5)~\text{nm}$ (inset \textbf{A}), whereas for high current density $j_e=1.34\cdot10^{12}~\text{A}/\text{m}^2$ the center position is pushed to the left hand side due to the spin-torque (inset \textbf{B}). 

So far the frequency of the domain wall oscillations can only be tuned by the injected current density, so that frequency and output power cannot be varied independently. By the application of an external magnetic field $\mathbf{H}_{ext}=H_{y}\mathbf{\hat{e}}_{y}$ in the plane of the wire, the potential landscape of the polar angle is modulated leading to fundamental changes in the power spectrum of the oscillations. The power spectra of the magnetization dynamics for various field strengths $\mu_{0}H_{y}= 0-3~\text{mT}$ and constant current density $j_{e}=1.79\cdot10^{11}~\text{A}/\text{m}^{2}$ (corresponding to spin drift velocity $u=1.48~\text{m/s}$) is plotted in Fig. 3. For $\mu_{0}H_{y}=0~\text{mT}$ it shows a sharp peak at the oscillation frequency $f=747~\text{MHz}$ whereas for higher field strength $\mu_{0}H_{y}> 2.0~\text{mT}$ the oscillation frequency strongly decreases continuously down to $f=525~\text{MHz}$, indicated by the red dashed line. For higher field strengths, additional peaks at higher frequencies appear. Note that in the graph, the first peak corresponding to the fundamental oscillation frequency is scaled down by a factor of $10$ for comparison.

In conclusion we have shown that Walker precession of a pinned domain wall can be easily obtained in perpendicularly magnetized materials, where the domain wall is pinned by the geometrical constriction of the wire. By properly choosing the wire constriction dimensions, a small $K_d$ can be attained, while keeping the domain wall strongly pinned. This combined with the high damping $\alpha$ results in Walker precession at low current density. This STO can be tuned over a broad frequency range by the injected current density.  When an external field is applied, the power spectrum is modified leading to strongly nonharmonic oscillations opening a novel way for tuning the frequency of the STO.

Finally we like to mention, that for a realistic device a high output power is a key criterion. This can be achieved by the coupling of multiple STO's and by signal enhancement through a magnetic tunnel junction, fabricated in top of the wire constriction yielding a three terminal device. The latter also opens an additional way of manipulating the oscillation frequency: The current in the plane of the wire, that excites the oscillation and sets the frequency, can be independently tuned from the current, that flows vertically across the tunnel barrier and determines the output power, which leads to a versatile microwave source.

\vspace{1cm}

The authors would like to acknowledge the financial support by the DFG (SFB 767, KL1811), the Landesstiftung Baden W\"urttemberg, the European Research Council via its Starting Independent Researcher Grant (ERC-2007-Stg 208162) scheme, the EU (SPINSWITCH MRTN-CT-2006-035327), and the Samsung Advanced Institute of Technology.


\clearpage

\vspace{1cm}
\begin{figure}[tp]
\includegraphics[width=10cm]{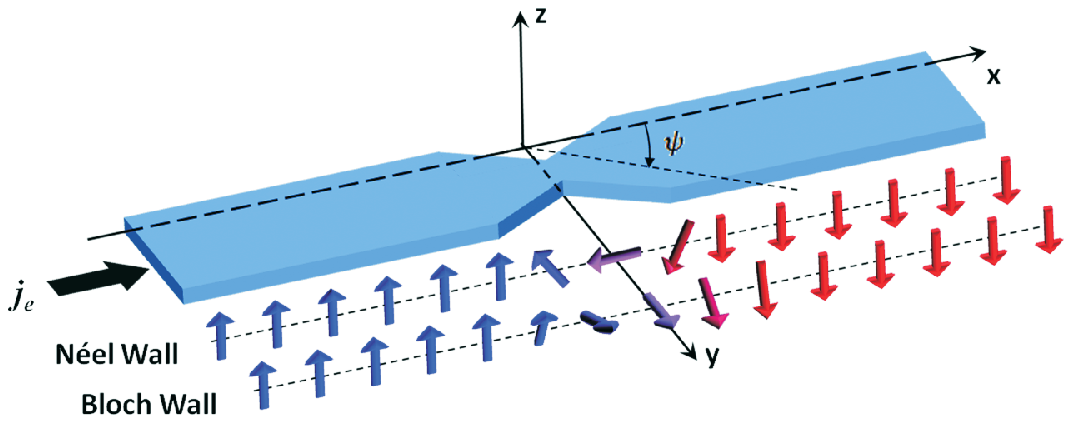}
\caption{Schematic illustration of the geometrically confined structure. The arrows represent the magnetization configuration inside the structure, which can be either a Bloch or a N\'eel domain wall.}
\label{fig1}
\end{figure}

\vspace{1cm}
\vspace{1cm}
\begin{figure}[tp]
\includegraphics[width=10cm]{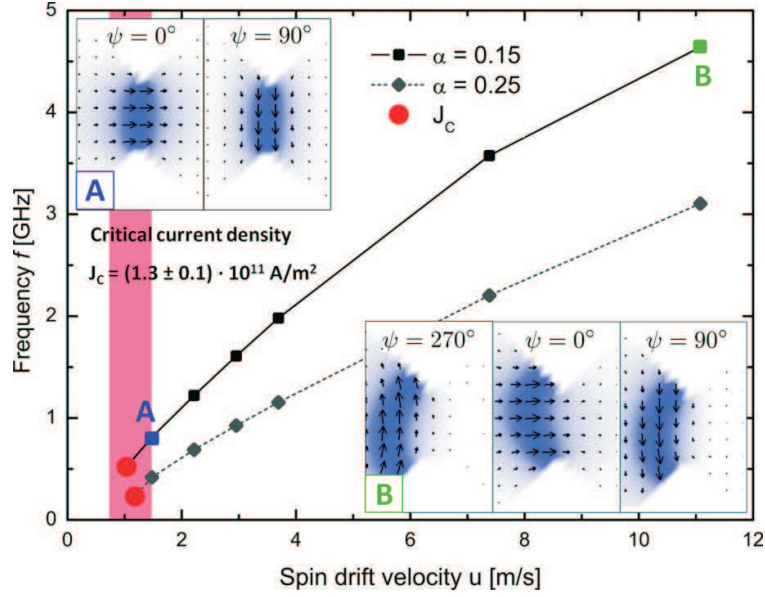}
\caption{Domain wall oscillation frequency $f$ as a function of injected spin-current drift velocity $u$ for constant $\alpha=(0.15, 0.25)$. The oscillation frequency shows a linear dependence on the current density over a broad frequency range between $500~\text{MHz}$ and $3.5~\text{GHz}$. (A) The domain wall profile is symmetric under rotation for low current density $j_e=1.79\cdot10^{11}~\text{A}/\text{m}^2$. (B) A high current density $j_e=1.34\cdot10^{12}~\text{A}/\text{m}^2$ the domain wall shows asymmetric oscillations, this leads to nonharmonic behaviour.}
\label{fig2}
\end{figure}

\begin{figure}[tp]
\includegraphics[width=10cm]{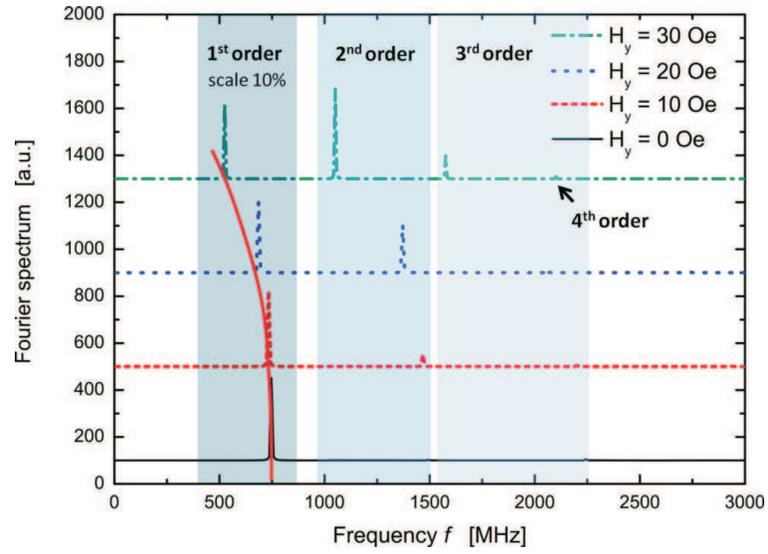}
\caption{Fourier spectrum of the magnetization component $M_{x}$ as a function of frequency $f$, for $j_{e}=1.79\cdot10^{11}~\text{A}/\text{m}^{2}$ and various external field strength $\mu_{0}H_{y} = (0~\text{mT},10~\text{mT}, 20~\text{mT}, 30~\text{mT})$. At zero field, a sharp peak indicates an oscillation frequency of $f=747~\text{MHz}$, whereas for higher field strenghts the oscillation frquency is shifted to lower frequencies down to $f=525~\text{MHz}$ (indicated by the red dashed line). Furthermore higher order oscillations are visible by the peaks forming at higher frequencies}
\label{fig3}
\end{figure}

\end{document}